\documentstyle[multicol,aps,prl,epsf]{revtex}
\begin{document}
\title{\bf Persistence of a Continuous Stochastic Process with Discrete-Time
Sampling}

\author{Satya N. Majumdar$^{(1),(2)}$, Alan J. Bray$^{(3)}$, and
George C. M. A. Ehrhardt$^{(3)}$}

\address{(1)Laboratoire de Physique Quantique (UMR C5626 du CNRS),
Universit\'e Paul Sabatier, 31062 Toulouse Cedex, France. \\
(2) Tata Institute of Fundamental Research, Homi Bhabha Road,
Mumbai-400005, India. \\
(3)Department of Physics and Astronomy, University of Manchester,
Manchester, M13 9PL, UK}

\date{\today}

\maketitle

\begin{abstract}

\noindent  We introduce  the concept  of  `discrete-time persistence',
which deals  with zero-crossings  of a continuous  stochastic process,
$X(T)$,  measured at discrete  times, $T=n\Delta  T$.  For  a Gaussian
Markov  process   with  relaxation  rate  $\mu$,  we   show  that  the
persistence  (no  crossing) probability  decays  as $[\rho(a)]^n$  for
large $n$, where $a=\exp(-\mu \Delta  T)$, and we compute $\rho(a)$ to
high  precision.    We  also   define  the  concept   of  `alternating
persistence', which corresponds to  $a<0$. For $a>1$, corresponding to
motion  in  an  unstable  potential  ($\mu<0$),  there  is  a  nonzero
probability of having no zero-crossings  in infinite time, and we show
how to calculate it.

\end{abstract}

\begin{multicols}{2}

Persistence  of a  continuous  stochastic process  has generated  much
recent interest in a  wide variety of nonequilibrium systems including
various  models  of phase  ordering  kinetics, diffusion,  fluctuating
interfaces and reaction-diffusion processes\cite{review}.  Persistence
has   also   been   recently    used   in   fields   as   diverse   as
ecology\cite{ecology}  and  seismology\cite{seismic}.  Persistence  is
simply the  probability $P(t)$ that  a stochastic process  $x(t)$ does
not change  sign up  to time  $t$.  In most  of the  systems mentioned
above,  $P(t)\sim t^{-\theta}$  for large  $t$, where  the persistence
exponent  $\theta$ is  nontrivial. Apart  from various  analytical and
numerical results, this exponent has also been measured experimentally
in    systems   such    as    breath   figures\cite{marcos},    liquid
crystals\cite{yurke},  soap bubbles\cite{tam},  and  more recently  in
laser-polarized Xe gas using NMR techniques\cite{wong}.

Persistence  has  also  remained   a  popular  subject  among  applied
mathematicians for many decades\cite{BL}.  They are most interested in
the probability of  `no zero crossing' of a  Gaussian {\it stationary}
process (GSP) between times $T_1$ and $T_2$\cite{slepian}.  It is well
known that this probability usually  decays as $\sim \exp (-\theta T)$
for large $T=|T_2-T_1|$ where $\theta$ is nontrivial\cite{slepian,BL}.
The  persistence  of  some   of  the  {\it  non-stationary}  processes
mentioned in  the previous paragraph such as  the diffusion processes,
can  be mapped to  that of  a corresponding  GSP\cite{diffusion}. This
makes the two sets of problems related to each other and the power law
exponent in the  former problem becomes the inverse  decay rate in the
latter.   Even  though  $\theta$  is,  in  general,  hard  to  compute
analytically, it is  very easy to evaluate numerically  in most cases.
Given  this  fact,  and  the  combined interest  of  both  statistical
physicists  and applied  mathematicians, much  recent effort  has been
devoted to computing $\theta$ numerically to extremely high precision.

This  raises  a  natural  question:  How accurately  can  one  measure
$\theta$? Is there a natural limitation and if so, can it be overcome?
This  issue arises  from the  following simple  observation.   All the
stochastic  processes  mentioned   above  occur  in  continuous  time.
However,  when one  performs numerical  simulations or  experiments on
persistence, one  has to  discretize time in  some way and  sample the
data only  at these discrete  time points to  check if the  process has
retained its  sign.  Due to  this discretization, some  information is
lost. For example, the process may have crossed and recrossed zero (or
a  spin may  have flipped  sign  many times)  between two  consecutive
discrete time points. These crossings  (or sign flips)  go  undetected
due to the discrete sampling of  the data. The question is how serious
is   this  loss   of  information.    Is  it   possible   to  estimate
quantitatively the error involved due to the discretization?

The purpose of this Letter is  twofold: (i) to point out that there is
indeed a very general and nontrivial effect, due to the discretization
of  time, on  the measured  persistence of  any  continuous stochastic
process, and (ii) to provide  a quantitative estimate of its magnitude
in a simple  Markov model. The effect turns out  to be nontrivial even
for  this  simple toy  model.   We  also  develop two  new  analytical
approaches,  perturbative and  variational, which  provide  results to
extremely high precision.  We  emphasize that, even though we restrict
ourselves here to a simple model  by way of an example, this effect is
very general and should be observable in simulations or experiments on
more realistic systems.

To  formulate  a precise  quantitative  question,  let  us consider  a
stationary stochastic process in  continuous time $T$ which is sampled
at times $T_1$, $T_2$, $\ldots$, $T_n=T$ separated by a uniform window
size, $T_i-T_{i-1}=\Delta T$ such  that $T=n\Delta T$.  The continuous
persistence  $P(T)$ is  then approximated  as $P(T)\approx  P_n$ where
$P_n$ is  the probability that the  process $X(T)$ is  positive at all
the $n$  discrete points. Note that,  for finite $\Delta  T$, $P_n$ is
different from $P(T)$ since the  process can cross zero more than once
between  two   successive  discrete  times.   One   expects  that  the
approximation  $P(T) \approx  P_n$ will  improve as  the  window size,
$\Delta T$, decreases, and in the limit $\Delta T\to 0$, $n\to \infty$
keeping $T=n\Delta T$ fixed, $P_n\to P(T)$. By contrast, if the window
size $\Delta T  \gg \tau$ where $\tau$ is the  correlation time of the
process, the stochastic variables  at different discrete points become
completely  uncorrelated and  we expect  $P_n \to  2^{-n}$,  since the
probability that at each point  the process is positive is just $1/2$.
We  then ask:  How  does the  discrete  persistence $P_n$  interpolate
between these two limits as $\Delta T$ varies continuously from $0$ to
$\infty$? We  show that for  a GSP, in general,  $P_n\sim [\rho(\Delta
T)]^n$  for  large  $n$,   where  the  function  $\rho(\Delta  T)$  is
nontrivial with the limiting behavior
\begin{equation}
\rho(\Delta  T)\approx\cases{ 1-\theta  \Delta  T, \,  &$\Delta T  \to
0$\cr 1/2, \, &$\Delta T \to \infty$, \cr}
\label{asym}
\end{equation}
where $\theta$ is the usual persistence exponent.  As $\Delta T\to 0$,
one  recovers  the continuous  persistence,  $P_n\to (1-\theta  \Delta
T)^n\sim \exp(-\theta T)$ where  $T=n\Delta T$. The general goal would
be to compute  this function $\rho(\Delta T)$, the  knowledge of which
will provide an  estimate of the difference, due  to the finite window
size $\Delta T$, between the measured persistence $P_n$ and the $P(T)$
of the underlying continuous process.

The  nonstationary  processes discussed  in  the  first paragraph  are
related  to   the  equivalent  stationary   ones  via  $T  =   \ln  t$
\cite{diffusion}.  A uniform spacing, $\Delta T$, between measurements
in  the   latter  systems,  therefore,   corresponds  to  measurements
uniformly spaced in  {\em log time} in the  former. Such a measurement
regime  has  indeed  been  used  in a  recent  experimental  study  of
diffusive  persistence   \cite{wong},  with  a   spacing  in  log-time
equivalent to $\Delta T \approx  0.24$. The present paper is the first
step  in understanding  how such  discretization affects  the measured
result. To compare directly with the experiment, we need
to compute the function $\rho(\Delta T)$ for the diffusion equation
which is hard due to the non-Markovian nature of the process.
However, to understand
the general nature of this function $\rho(\Delta T)$, it would be
useful to find a toy model where it can be computed explicitly.
We consider  below a simple Gaussian Markov
process for which progress can  be made in that direction.
The physical process we study
is   the  one-dimensional  Ornstein-Uhlenbeck   motion  of   a  noisy,
overdamped  particle  in a  potential  $V(X)={\mu  X^2}/2$, where  the
position $X$ of the particle evolves via the Langevin equation,
\begin{equation}
{{d X}\over {dT}}=-\mu X + \eta(T).
\label{lange}
\end{equation}
The  white noise  $\eta(T)$ has  zero mean  and a  correlator $\langle
\eta(T) \eta(T')\rangle = 2D\delta(T-T')$.

For  this process, we  first evaluate  the continuous  persistence and
then  compute  the  function  $\rho(\Delta T)$.   For  the  continuous
persistence, a  backward Fokker-Planck  (BFP) approach is  useful. Let
$Q(X,T)$ denote  the probability that,  starting at $X$ at  $T=0$, the
particle has not crossed the origin, $X=0$, up to time $T$.  We expect
different behavior depending on  whether $\mu>0$ (stable potential) or
$\mu<0$ (unstable  potential).  In the former case,  the particle will
eventually   cross  the   origin   and  hence   $Q(X,T)$  will   decay
exponentially with time. In the latter case, however, the particle has
a  finite probability  to escape  to infinity,  and  hence persistence
should decay to  a nonzero number. The latter case  is also related to
the problems of escape from metastable states studied before\cite{ms}.

The probability $Q(X,T)$ satisfies the BFP equation,
\begin{equation}
{  {\partial Q}\over  {\partial T}}=D  {{\partial^2  Q}\over {\partial
X^2}}-\mu X {{\partial Q}\over {\partial X}},
\label{cbfp}
\end{equation}
with boundary  conditions $Q(0,T)=0$ and $Q(\infty,T)=1$  for all $T$,
and initial condition $Q(X,0)=1$ for all $X>0$. The solution is
\begin{equation}
Q(X,T)= {\rm  Erf} \left[ { {e^{-\mu T}}  \over {\sqrt {2D'(1-e^{-2\mu
T})} }} X\right],
\label{mup}
\end{equation}
where $D'=D/\mu$ and ${\rm Erf}[x]$ is the error function.  For $\mu >
0$,  $Q(X,T)$  becomes  separable  in  $X$  and  $T$  for  large  $T$,
$Q(X,T)\sim e^{-\mu T}X$, and  decays exponentially with $T$ for fixed
$X$.  This gives the persistence exponent $\theta=\mu$. For $\mu < 0$,
however,  $Q(X,T)$ approaches  the steady  state solution  $Q(X)= {\rm
Erf}\,(X/\sqrt{2|D'|})$  as $T\to  \infty$.   We also  note from  Eq.\
(\ref{lange}) that  the critical case $\mu=0$  corresponds to ordinary
Brownian motion, and taking the  limit $\mu\to 0$ in Eq.\ (\ref{mup}),
one  recovers  the known  result,  $Q(X,T) ={\rm  Erf}[X/\sqrt{4DT}]$,
which decays as a power law, $Q(X,T)\sim X/{\sqrt T}$, for large $T$.

For  later   purposes,  we  will   also  need  the   Green's  function
$G(X_2,T_2|X_1,T_1)$,  the probability that  the particle  starting at
$X=X_1$ at  $T=T_1$ will  reach $X_2$ at  $T_2$, with  $T_2>T_1$. This
propagator can be easily  computed exactly from Eq.\ (\ref{lange}) and
we get,
\begin{equation}
G(X_2,T_2|X_1,T_1)    =    \frac{1}{\sqrt{2\pi   D'    (1-a^2)}}\,e^{-
\frac{(X_2-a X_1)^2}{2D'(1-a^2)}},
\label{green}
\end{equation}
where $a=e^{-\mu(T_2-T_1)}$.  Note that  for $\mu \ge 0$, $0\leq a\leq
1$, while for $\mu<0$, $a > 1$ (and $D' = D/\mu<0$).

We  now turn  to  the  discrete persistence  $P_n$  of the  continuous
process in  Eq.\ (\ref{lange}). Let  $Q_n(X)$ be the  probability that
starting at $X$ at $T=0$, the  process is positive at all the discrete
points  $T_1=\Delta  T$, $T_2=2\Delta  T$,  $\ldots$, $T_n=n\Delta  T$
separated by  the uniform  window size $\Delta  T$. Then  the discrete
persistence is  $P_n=\int_0^{\infty}Q_n(X)P_0(X)dX$, where $P_0(X)$ is
the distribution  of the initial position  of the particle  and can be
arbitrary.   Using  the  Markov   property  of  the  process  in  Eq.\
(\ref{lange}),  it is  easy to  write down  a recurrence  relation for
$Q_n(X)$,
\begin{equation}
Q_{n+1}(X)=\int_0^{\infty}G(Y,\Delta T|X,0)\,Q_n(Y)dY,
\label{rec1}
\end{equation}
where $G$ is  the propagator as in Eq.\  (\ref{green}) with $a=e^{-\mu
\Delta  T}$ and  $Q_0(X)=1$ for  all  $X>0$.  This  recurrence is  the
discrete analogue of the continuous BFP equation (\ref{cbfp}). Indeed,
it can be checked that  Eq.\ (\ref{rec1}) reduces to Eq.\ (\ref{cbfp})
in the limit $\Delta T\to 0$. To simplify the algebra, we consider the
rescaled variable,  $x=X/{\sqrt {D'(1-a^2)}}$,  in terms of  which the
recursion reads
\begin{equation}
Q_{n+1}(x)=    {1\over     {\sqrt    {2\pi}}}\int_0^{\infty}\exp    [-
(y-ax)^2/2]Q_n(y)dy,
\label{rec2}
\end{equation}
where  we  have  used  the  explicit  expression  for  $G$  from  Eq.\
(\ref{green}).

Let us first consider the  case $\mu>0$, i.e., $0\leq a=e^{-\mu \Delta
T}<1$,  where, guided  by the  continuous case,  we  expect $Q_n(x)\to
{\rho}^n q(x)$ as  $n\to \infty$ at any fixed  $x$.  Substituting this
asymptotic form into Eq.\  (\ref{rec2}), we get an integral-eigenvalue
equation for $q(x)$,
\begin{equation}
\rho       q(x)={1\over      {\sqrt      {2\pi}}}\int_0^{\infty}\exp[-
(y-ax)^2/2]q(y)dy,
\label{eigen}
\end{equation}
with eigenvalue $\rho(a)$ that  evidently depends continuously on $a$.
Although Eq.\ (\ref{eigen}) admits many eigenvalues, we are interested
only  in the  largest  eigenvalue since  it  dominates the  asymptotic
behavior  of  $Q_n(x)$  for  large   $n$.   We  also  note  that  Eq.\
(\ref{eigen})  determines  the  eigenfunction  $q(x)$ only  up  to  an
overall  multiplicative constant.   Let  us first  consider the  limit
$a\to 0$  or equivalently $\Delta T  \to \infty$.  In  this case, Eq.\
(\ref{eigen}) can be solved exactly to give $\rho=1/2$ and $q(x)= {\rm
const}$,  thus recovering  the correct  limiting  behavior, $Q_n(x)\to
{\rm const}\ 2^{-n}$ (and the initial condition, $Q_0(x)=1$ for $x>0$,
fixes  the constant  at  unity).   For small  $a$,  by expanding  Eq.\
(\ref{eigen})  in a  Taylor series,  it is  easy to  compute $\rho(a)$
perturbatively, giving $\rho = {1\over {2}}+ {1\over {\pi}}a +O(a^2)$.
The goal now is to evaluate  $\rho(a)$ for arbitrary $a$.  To this end
we develop below  two analytical approaches and compare  them with the
direct numerical integration of Eq. (\ref{rec2}).

{\it Perturbative  approach}: We  expand the factor  $\exp(axy)$, from
the exponential in Eq.\ (\ref{eigen}), as a power series and integrate
term by term, to get
\begin{eqnarray}
\label{matrix1}
\rho  q(x) & =  & \frac{\exp(-a^2x^2/2)}{\sqrt{2\pi}}\sum_{n=0}^\infty
\frac{b_n}{\sqrt{n!}}\,(\sqrt{a}x)^n,     \\     b_n     &     =     &
\frac{a^{n/2}}{\sqrt{n!}}\int_0^\infty dy\, y^n\exp(-y^2/2)\,q(y).
\label{matrix2}
\end{eqnarray}
Substituting (\ref{matrix1}) into  (\ref{matrix2}) leads to the matrix
eigenvalue equation
\begin{eqnarray}
\label{matrix3}
\rho  b_n  &  = &  \sum_{m=0}^\infty  A_{nm}\,b_m,  \\  A_{nm} &  =  &
\frac{1}{\sqrt{4\pi(1+a^2)}}    \left(\frac{2a}{1+a^2}\right)^{(n+m)/2}
\frac{\Gamma(\frac{n+m+1}{2})}{\sqrt{n!m!}}.
\label{matrix4}
\end{eqnarray}
This approach  converts an integral eigenvalue equation  into a matrix
eigenvalue equation, with  matrix elements that decrease exponentially
as $n$  and $m$ increase. Computing  the largest eigenvalue  of the $N
\times N$ submatrix  ($n,m=0,1,\ldots,N-1$) gives a rapidly converging
series of estimates for $\rho$ as  $N$ increases. For a given $N$, the
result     is    exact     to     order    $\epsilon^{N-1}$,     where
$\epsilon=2a/(1+a^2)$.  In this way  one can easily obtain results for
$\rho(a)$  correct  to one  part  in  $10^{12}$.  Convergence  becomes
progressively slower as  $a \to 1$, which is  expected since $\epsilon
\to 1$ in this limit.  For  $a \to 1$, however, we have the analytical
result  $\rho \to a$  [such that  $\rho^n \to  \exp(- n\mu\Delta  T) =
\exp(-\mu T)$],  since we  must recover the  continuum result  in this
limit.

%Note, finally, that from the eigenvector, $\{b_n\}$, associated
%with the largest eigenvalue, we can reconstruct, using
%(\ref{matrix1}), the corresponding eigenfunction $q(x)$.

{\it  Variational  approach}:  It  is  possible  to  derive  a  useful
variational inequality  for $\rho$.  First  we note that  the integral
operator in Eq.\ (\ref{eigen}), asymmetric in $x$ and $y$, can be made
self-adjoint  via   the  substitution,  $q(x)=g(x)\exp  [{(1-a^2)\over
{4}}x^2]$ which gives,
\begin{equation}
\rho g(x)={1\over {\sqrt {2\pi}}}\int_0^{\infty} K(x,y)g(y)dy,
\label{selfad}
\end{equation}
where  $K(x,y)=K(y,x)=\exp  [-{(1+a^2)\over {4}}(x^2+y^2)+axy]$.   Let
$f(x)$  be  any  normalizable  function,  $\int_0^{\infty}f^2(x)dx=1$.
Using  elementary   properties  of   linear  vector  spaces   and  the
self-adjoint  property of  the integral  operator, it  becomes evident
from Eq.\ (\ref{selfad}) that  the largest eigenvalue $\rho$ satisfies
the inequality,
\begin{equation}
\rho          \geq           {1\over          {\sqrt          {2\pi}}}
\int_0^{\infty}\int_0^{\infty}f(x)K(x,y)f(y) dx dy.
\label{ineq}
\end{equation}
One  can then use  any trial  function $f(x)$  containing one  or more
variational parameters and  then maximize the right hand  side of Eq.\
(\ref{ineq})  with respect to  these parameters  to derive  a rigorous
lower bound for $\rho(a)$ for arbitrary $0<a<1$.

The  limiting   forms  of  the  true  eigenfunction   $g(x)$  in  Eq.\
(\ref{selfad}) for $a\to 0$ and $a\to 1$ can be easily worked out, and
suggest  a trial  function  of the  form  $f(x)=A (b+x)\exp  (-\lambda
x^2/2)$. The  amplitude $A$ is  fixed by the  normalization condition,
$\int_0^{\infty}f^2(x)dx=1$,  while  $b$  and  $\lambda$ are  the  two
variational parameters.  The right-hand side of the inequality in Eq.\
(\ref{ineq}) can then be evaluated in closed form and the optimization
with  respect   to  $b$   and  $\lambda$  performed.    The  resulting
variational estimate  turns out to  be very accurate for  all $0<a<1$,
when compared  to numerical results, and agrees  with the perturbative
results to at least 4 or 5 decimal places.

{\it  Numerical Integration}: It  is not  difficult to  integrate Eq.\
(\ref{rec2}) directly. However, since  $Q_n(x)\to 1$ as $x\to \infty$,
numerically  it  is  convenient   to  first  make  the  transformation
$Q_n(x)=G_n(x)\exp[(1-a^2)x^2/4]$ in Eq.\  (\ref{rec2}) and then study
the resulting  equation for $G_n(x)$  by numerical iteration,  with an
arbitrary  initial condition.   For large  $n$, $G_n(x)$  converges to
${\rho}^n g(x)$  where $g(x)$ is the solution  of Eq.\ (\ref{selfad}).
The eigenvalue $\rho$  is determined from the slope  of the log-linear
plot  of $A_n=\int_0^{\infty}G_n(x)dx \sim  {\rho}^n$ versus  $n$.  In
Table  1,  we compare  the  numerical,  variational, and  perturbative
estimates of $\rho$.  The differences  are small in all cases, and the
variational bound is satisfied.

The eigenfunction $q(x)$ of  Eq.\ (\ref{eigen}) can also be calculated
by using  the series (\ref{matrix1}), with  the coefficients $\{b_n\}$
obtained from  the corresponding eigenvector  of the matrix  $A$, Eq.\
(\ref{matrix4}).   It is  shown, for  $a=0.5$, as  the lower  curve in
Fig.\  1.  The  asymptotic large-$x$  behavior (dashed  curve)  can be
obtained  analytically by noting  that for  large $x$  we can  set the
lower limit in (\ref{eigen})  to minus infinity with negligible error.
The resulting equation can  be solved exactly \cite{GR}, with solution
$q(x) = \exp(X^2/4)D_\nu(X)$,  where $X=(1-a^2)^{1/2}x$, $D_\nu(X)$ is
the  parabolic cylinder  function, and  $\nu =  \ln \rho/\ln  a$.  The
asymptotic  behavior  is $q(x)  \sim  x^\nu$.   The variational  trial
function, however,  misses this asymptotic behavior (see  Fig. 1) even
though the variational eigenvalue is very accurate.

\begin{center}
\begin{tabular}{||l||l||l||l||} \hline
 $a$ & $\rho_{num}$ & $\rho_{var}$ & $\rho_{pert}$ \\ \hline 1.0 & 1 &
1 &1 \\ 0.8 & 0.8524547  & 0.852440 &0.852454696506 \\ 0.6 & 0.7405959
&   0.740589   &0.740595939159  \\   0.4   &   0.6477666  &   0.647765
&0.647766585747 \\ 0.2 & 0.5684903 & 0.568490 &0.568490321623 \\ 0.0 &
1/2 & 1/2 & 1/2 \\ -0.2 & 0.4408132 & 0.440813 &0.440813209205 \\ -0.4
& 0.3900580 & 0.390004 &0.390057988652  \\ -0.6 & 0.3469679 & 0.346814
&0.346967773049 \\ -0.8 & 0.3106439 & 0.310444 &0.310643770245 \\ -1.0
& 0.2800859 & 0.279890 &0.280085758710 \\ \hline
\end{tabular}
\end{center}

\noindent Table1. Estimates of the  eigenvalue $\rho(a)$ for $-1 \le a
\le  1$, from  numerical, variational  and perturbative  methods.  The
latter is  the most precise, being  accurate to the  number of figures
quoted.
\medskip

Although Eq.\  (\ref{rec2}) was  derived for $a\geq  0$, one  can also
study  this equation  or,  equivalently, Eq.\  (\ref{eigen}) and  Eq.\
(\ref{selfad}),  for negative $a$.   Is there  a physical  meaning for
negative $a$?  Let $R_n(x)$ denote discrete `alternating' persistence,
being the probability  that, starting at $x>0$ ($x$  is related to $X$
as before) at $T=0$, the particle's position changes sign at alternate
discrete points up  to the $n$-th step. Then  $R_n(x)$ evolves via the
recurrence equation,
\begin{equation}
R_{n+1}(x)={1\over {\sqrt
{2\pi}}}\int_{-\infty}^{0}\exp[-(y-ax)^2/2]R_n(y)dy.
\label{rec3}
\end{equation}
Changing  $y\to -y$  inside the  integral, and  using $R_n(y)=R_n(-y)$
(since the process  has zero mean), we find  Eq.\ (\ref{rec3}) reduces
to   Eq.\   (\ref{rec2})   with   $a$   replaced   by   $-a$.    Thus,
$R_n(x,a)=Q_n(x,-a)$  and hence the  largest eigenvalue  $\rho(a)$ for
negative $a$  governs the  asymptotic decay of  `alternating' discrete
persistence.   We  also  note   that  while,  for  $a>0$,  $Q_n(x)\sim
\rho^n(a)q(x)$  for large  $n$  only for  $a<1$  (for $a>1$,  $Q_n(x)$
approaches a steady state -- see later), for negative $a$, $Q_n(x)\sim
\rho^n(a)q(x)$ for all  $a<0$. Furthermore, from Eq.\ (\ref{matrix4}),
one has the  symmetry relation $\rho(1/a) = |a|\rho(a)$,  which can be
used to obtain $\rho$ (and the corresponding eigenfunction) for $a<-1$
from the results for $-1 <a  \le 0$. In particular, $\rho \to 1/2$ for
$a  \to  0$  implies  $\rho  \to  1/2|a|$ for  $a  \to  -\infty$.   

Finally, we  turn to the unstable potential,  $\mu<0$, i.e. $a=e^{-\mu
\Delta T}>1$.  As in the  continuous case, we expect that the solution
of Eq.\  (\ref{rec2}) for  $a>1$ will reach  a steady state  for large
$n$, $Q_n(x)\to  q(x)$, where $q(x)$ will  satisfy Eq.\ (\ref{eigen}),
but with  $\rho=1$. Evidently  $q(x)$ will depend  on $a$, and  in the
limit  $a\to  1^{+}$  (i.e.   $\Delta  T\to  0$)  it  reduces  to  the
continuous result obtained from  Eq.\ (\ref{mup}).  For general $a>1$,
it is again possible to  obtain accurate variational and very accurate
perturbative estimates for $q(x)$. We omit the details here since they
are  somewhat similar  to the  $a<1$ case.   In Fig.\  1, we  plot the
perturbative $q(x)$ for $a=2$  (upper curve).  The variational result,
and  the  numerical result  obtained  from  direct  iteration of  Eq.\
(\ref{rec2}), are both indistinguishable  from the plotted curve. Note
that the case $a<-1$, discussed in the previous paragraph, corresponds
to alternating  persistence in an unstable potential,  which does {\em
not} approach a steady state.
\begin{figure}
\narrowtext\centerline{\epsfxsize\columnwidth
\epsfbox{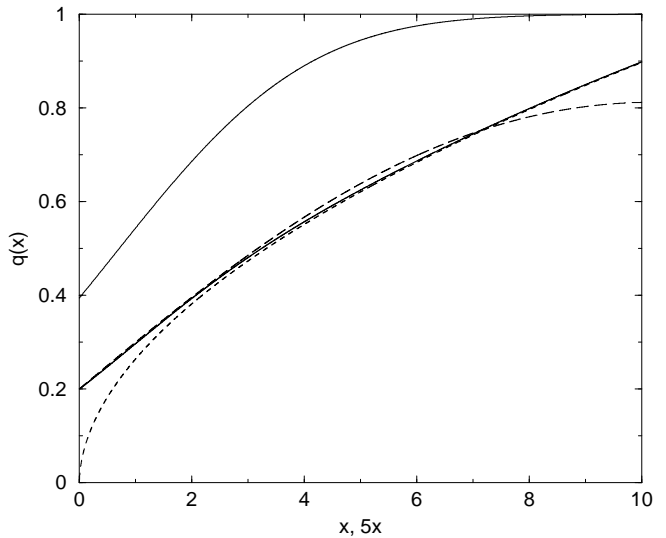}}\caption{The eigenfunctions $q(x)$ for $a=0.5$
(lower curve) and $a=2.0$  (upper curve, abscissa $=5x$).  Solid lines
- perturbative results; long-dashed - variational; dashed - asymptotic
result $q(x) \sim x^\nu$, with $\nu=\ln\rho/\ln a \simeq 0.530661$ for
$a=0.5$.}
\label{wf}
\end{figure}

In summary,  we have  shown that the  discrete persistence due  to the
finite  size  of  the  time  windows  differs  considerably  from  the
continuous persistence usually studied and we have computed explicitly
this  nontrivial effect analytically  for a  simple Markov  model. The
work extending some of the techniques developed here to more realistic
non-Markov processes is in progress.  We conclude by noting the recent
examples of discrete time persistence in dynamical systems\cite{bp}.

\end{multicols}

\end{document}